\documentclass[12pt]{article}
\usepackage[symbol]{footmisc}
\def\correspondingauthor{\footnote{Corresponding author.  }}
\usepackage[left=2cm,top=2.5cm,right=2cm,bottom=2.5cm]{geometry}
\usepackage{amsmath, amssymb}
\DeclareMathOperator{\sech}{sech}
\usepackage{graphicx}
\usepackage{graphics}
\usepackage{epstopdf}
\usepackage{subfigure}
\usepackage{amsfonts}
\usepackage{sectsty}
\usepackage{sectsty}
\usepackage{hyperref}
\usepackage{cite, multirow,hhline}

\begin{document}
	\begin{center}
	\large{\bf{Non violation of energy conditions in wormholes modelling}} \\
	\vspace{5mm}
	\normalsize{Nisha Godani$^1$ and Gauranga C. Samanta$^{2,}{}$\correspondingauthor{}, }\\
	\normalsize{$^1$ Department of Mathematics, Institute of Applied Sciences and Humanities\\ GLA University, Mathura, Uttar Pradesh, India\\
	$^2$ Department of Mathematics, BITS Pilani K K Birla Goa Campus, Goa, India}\\
	\normalsize {nishagodani.dei@gmail.com \\gauranga81@gmail.com}
\end{center}
	\begin{abstract}
	Morris \& Thorne \cite{morris1} proposed geometrical objects called traversable wormholes that act as bridges in connecting two spacetimes or two different points of the same spacetime. The geometrical properties of these wormholes depend upon the choice of the shape function. In literature, these are studied in modified gravities for different types of shape functions. In this paper, the traversable wormholes having shape function $b(r)=\frac{r_0\tanh(r)}{\tanh(r_0)}$ are explored in $f(R)$ gravity with $f(R)=R+\alpha R^m-\beta R^{-n}$, where $\alpha$, $\beta$, $m$ and $n$ are real constants. For different values of constants in function $f(R)$, the analysis is done in various cases. In each case, the energy conditions, equation of state parameter and anisotropic parameter are determined.
\end{abstract}

\textbf{Keywords:} Traversable Wormhole; $f(R)$ gravity; Energy condition
\section{Introduction}
Wormholes are hypothetical structures in space-time that connects two distinct universes is called inter-universe wormholes or two different regions of the same space-time in the same universe is called intra-universe wormholes. The difference between these two classes of wormhole ascends only at
the global level of geometry and global level of topology. Near the throat of the wormholes local physics is oblivious to issues of intra-universal or inter-universal travel.     Flamm \cite{flamm}  first studied wormhole type solutions. Then Einstein-Rosen carried out the study of these geometrical objects and obtained a solution which is known as Einstein-Rosen bridge \cite{eins-ros}. The word wormhole was first introduced by Misner and Wheeler \cite{misner}. They showed that wormholes cannot be traversable for standard
matter due to its instability.  Ellis \cite{ellis} and Bronnikov \cite{bronnikov} first assumed an example of wormhole belonging to the Morris Thorne class and the corresponding solution is known as Ellis-Bronnikov wormhole. Later on,  this wormhole was found to be instable \cite{Shinkai, Gonzalez, Gonzalez1, Doroshkevich, Bronnikov,  Bronnikov1}. The interest in the study of wormholes started with the work of Morris and Thorne \cite{morris1}. They provided some conditions for a wormhole to be traversable. They showed that the presence of exotic matter, matter different from the normal matter, is necessary for the existence of traversable wormholes. The word ``traversable'' is used here to indicate that a human could travel through the wormhole within a reasonable amount of time. Therefore, event horizons should not be present in the system for the existence of traversable wormholes, if event horizon present, then it should be out of the way so that the traveler does not have to cross it. Morris et al. \cite{morris2} showed that the wormhole structure in space-time, developed for interstellar travel, can be transformed into time-machine. The exotic matter includes the violation of null energy condition. The stress energy distribution is certainly abnormal at the throat of a wormhole. In a wormhole, when a bunch of converging radial light rays enters, then it reaches at the wormhole throat and at this point it losses null convergence condition. It becomes parallel  at the throat of wormhole and then enters into the other side and diverges. Thus, the null convergence condition is not satisfied at throat \cite{visser} and hence the null energy condition is also not satisfied, in general relativity. Eventually, the matter near the throat of the wormhole does not satisfy the null energy condition. Also, some negative energy density may be observed, somewhere near the throat of the wormhole. As a result, other energy conditions are also violated \cite{hawking, wald}. But this violation of energy conditions is not necessary in modified theories of gravity because the function in gravitational action is different than the function in GR and hence the formed field equations are completely different. Therefore, this demands investigation of wormhole structure and null energy conditions in alternate theories.

Among various alternate theories developed in literature, $f(R)$ theory of gravity is one that generalizes the Einstien's theory of general relativity. It includes an arbitrary function $f(R)$ of  Ricci scalar $R$ in gravitational action in place of function equal to $R$.  This theory have been studied in various aspects \cite{star, sotiriou, huang, noj, cog2008, felice, Bamba4, Bamba3, Bamba1, sebas, thakur, Bamba, Peng, Motohashi, Astashenok, noj2017, baha1, Yousaf, Faraoni, Abbas, Sussman, Mongwane, Mansour, Muller, Wang, Papagiannopoulos, Oikonomou, Chakraborty, Nashed, Capozziello, Gu, Abbas1, Mishra, Odintsov, godani}. The exploration of wormhole solutions has become an important aspect of study and various cosmologists have paid their attention and provided a good utilization of $f(R)$ theory.
% 2003
Lemos et al. \cite{lemos} reviewed wormhole literature, analyzed Morris and Thorne metric in the presence of cosmological constant and studied properties of traversable wormholes.
%2006
Rahaman et al. \cite{rahaman} considered the cosmological constant as a function of radial coordinate, obtained wormhole solutions and determined a condition for averaged null energy condition to be small.
Lobo and Oliveira \cite{lobo} investigated the existence of wormholes in the framework of $f(R)$ theory of gravity.
%2010
Garcia and Lobo \cite{garcia} assumed coupling between a function of scalar curvature and Lagrangian density of matter and studied  wormholes in the context of modified gravity. They used linear and non-minimal coupling between curvature and matter and obtained an exact solution.
%2015
Duplessis and Easson \cite{dupe} used  pure, scale-free $R^2$ gravity and obtained exotic solutions of traversable wormhole and black hole.
%2016
Bahamonde et al. \cite{baha} analysed wormhole solutions in modified $(R)$ gravity. They used the equation of state for matter and radiation filled universes and obtained a wormhole structure approaching to FLRW universe asymptotically.
Wang and Meng \cite{wang} considering wormholes in bulk viscosity determined wormhole solutions, studied the  factors upon which the value of  traversal velocity depends and found the conditions for the violation of null energy condition.
%2017
Moradpour \cite{morad} studied traversable wormholes in general relativity. He also studied wormhole geometry in Lyra manifold and obtained some traversable wormholes which are asymptotically flat.
Zubair et al. \cite{zubair} used $f(R,\phi)$ gravity for the examination of symmetric, spherical and static  wormhole solutions. They considered three types of  fluids and analyzed the energy conditions. They obtained constraints for the validation of weak and null energy conditions and showed a possibility for the construction of realistic wormhole structures.
%2018
Shaikh \cite{shaikh} obtained wormhole solutions in Eddington-inspired Born-Infeld gravity and found the existence of the matter which is non-exotic in nature and satisfies the energy conditions.
Barros and Lobo \cite{barros} showed that the wormhole geometries can be supported with the use of three-form fields. They also found the validation of weak and null energy conditions due to presence of these fields.
Peter \cite{peter} used the framework of $f(R)$ gravity for the investigation of traversable wormholes. He determined several wormhole solutions with respect to various shape functions, then he also derived the $f(R)$ functions. He showed that the combine effect of non-commutative geometry and $f(R)$ gravity leads to violation of null energy condition. Many other cosmologists have also made their significant contribution in the study of wormholes \cite{saiedi, Zangeneh1, lopez, najafi, Rahaman, Mehdizadeha, Zangeneh, Mehdizadeh, Moraes, Bejarano, cataldo, Rogatko, Paul, novi, Ovgun, Tsukamoto}.
Recently, Godani and Samanta \cite{godani} studied traversable wormhole in $f(R)$ gravity with two different shape function. In this paper, we would like
to investigate the traversable wormhole in $f(R)$ gravity with hyperbolic shape function, where $f(R) = R + \alpha R^m - \beta R^{-n}$ ($\alpha, \beta$, $m$ and $n$ are real constants)\cite{noj1}. The motivation of this paper is to investigate the existence of the traversable wormhole without violation of energy conditions for different values of $\alpha, \beta$, $m$ and $n$.

%Sahoo et al. \cite{sahoo} used particular red shift and shape functions and obtained wormhole solutions in the setting $f(R,T)$ gravity. They investigated the energy conditions and found the wormholes to be filled with Phantom fluid.

The paper is organized in the  following manner. In Sec-2, the Einstein's field equations are given for static traversable wormholes in $f(R)$ gravity.
In Sec-3, wormhole solutions in $f(R)$ model are obtained with a hyperbolic   shape function. In Sec-4, the results are discussed. Finally, in Sec-5, the work is concluded.

\section{Traversable Wormhole \& Equations}
%In this section, $f(R)$ gravity and Einstein's field equations for FRW metric are described  briefly. Throughout dot and dash upon a function denote derivative with respect to cosmic time and Ricci scalar respectively.
The  metric defining the wormhole structure  is
  \begin{equation}\label{metric}
ds^2=-e^{2\Phi(r)}dt^2+\frac{dr^2}{1-b(r)/r} + r^2(d\theta^2+\sin^2\theta^2\phi^2),
\end{equation}
which is a spherically symmetric static metric. The range of the coordinate $r$ is $-\infty$ to $\infty$. The redshift function $\Phi (r)$ must be constant everywhere for traversable wormhole. The function $\Phi(r)$ is responsible for the determination of gravitational redshift, hence it is called redshift function. The two asymptotically flat regions are assumed to be occur at $r\approx\pm \infty$.
The wormhole solutions must satisfy Einstein's field equations and must possess a throat that joins two regions of universe which are asymptotically flat. For a traversable wormhole, event horizon should not be present and the effect of tidal gravitational forces should be very small on a traveler.

The functions $\Phi(r)$ and $b(r)$ are the functions of  radial coordinate $r$, which is a non-decreasing function. Its minimum value is $r_0$, radius of the throat, and maximum value is $+\infty$.  The function $b(r)$ is responsible for the shape of wormhole, hence it is known as shape function. The existence of wormhole solutions demands the satisfaction of following conditions:
(i) $b(r_0)=r_0$, (ii) $\frac{b(r)-b'(r)r}{b^2}>0$, (iii) $b'(r_0)-1\leq 0$, (iv) $\frac{b(r)}{r}<1$ for $r>r_0$ and (v) $\frac{b(r)}{r}\rightarrow 0$ as $r\rightarrow\infty$. For simplicity, the redshift function is assumed as a constant.

Morris \& Thorne \cite{morris1} introduced traversable wormholes using the background of Einstein's general theory of relativity. The $f(R)$  theory of gravity generalizes  Einstein's theory of relativity by replacing the  gravitational action $R$ with a general function $f(R)$ of $R$. Thus, this action for $f(R)$ theory  is given as
\begin{equation}\label{action}
S_G=\dfrac{1}{2k}\int[f(R) + L_m]\sqrt{-g}d^4x,
\end{equation}
where $k=8\pi G$, $L_m$ and $g$ stand for the  matter Lagrangian density and  the  determinant of the metric $g_{\mu\nu}$ respectively. For simplicity $k$ is taken as unity.\\

Variation of Eq.(\ref{action}) with respect to the metric $g_{\mu\nu}$ gives the field equations as
\begin{equation}\label{fe}
FR_{\mu\nu} -\dfrac{1}{2}fg_{\mu\nu}-\triangledown_\mu\triangledown_\nu F+\square Fg_{\mu\nu}= T_{\mu\nu}^m,
\end{equation}	
where $R_{\mu\nu}$ and $R$ denote Ricci tensor and curvature scalar respectively and $F=\frac{df}{dR}$. The contraction of \ref{fe}, gives
\begin{equation}\label{trace}
FR-2f+3\square F=T,
\end{equation}
where $T=T^{\mu}_{\mu}$ is the trace of the stress energy tensor.

From Eqs. \ref{fe} \& \ref{trace}, the effective field equation is obtained as
\begin{equation}
G_{\mu\nu}\equiv R_{\mu\nu}-\frac{1}{2}Rg_{\mu\nu}=T_{\mu\nu}^{eff},
\end{equation}
where $T_{\mu\nu}^{eff}=T_{\mu\nu}^{c}+T_{\mu\nu}^{m}/F$ and $T_{\mu\nu}^{c}=\frac{1}{F}[\triangledown_\mu\triangledown_\nu F-\frac{1}{4}g_{\mu\nu}(FR+\square F+T)]$.
The energy momentum tensor for the matter source of the wormholes is $T_{\mu\nu}=\frac{\partial L_m}{\partial g^{\mu\nu}}$, which is defined as
\begin{equation}
T_{\mu\nu} = (\rho + p_t)u_\mu u_\nu - p_tg_{\mu\nu}+(p_r-p_t)X_\mu X_\nu,
\end{equation}	
such that
\begin{equation}
u^{\mu}u_\mu=-1 \mbox{ and } X^{\mu}X_\mu=1,
\end{equation}

where $\rho$,  $p_t$ and $p_r$  stand for the energy density, tangential pressure and radial pressure respectively.

The  Ricci scalar $R$ given by $R=\frac{2b'(r)}{r^2}$ and Einstein's field equations for the metric \ref{metric} in  $f(R)$ gravity are obtained as:
%\cite{lobo}:
%\begin{eqnarray}  \label{6}
\begin{equation}\label{6}
\rho=\frac{Fb'(r)}{r^2}-H
\end{equation}
\begin{equation}\label{7}
p_r=-\frac{b(r)F}{r^3}-\Bigg(1-\frac{b(r)}{r}\Bigg)\Bigg[F''+\frac{F'(rb'(r)-b(r))}{2r^2\Big(1-\frac{b(r)}{r}\Big)}\Bigg]+H
\end{equation}
\begin{equation}\label{8}
p_t=\frac{F(b(r)-rb'(r))}{2r^3}-\frac{F'}{r}\Bigg(1-\frac{b(r)}{r}\Bigg)+H,
\end{equation}

where $H=\frac{1}{4}(FR+\square F+T)$ and prime upon a function denotes the derivative of that function with respect to  radial coordinate $r$.

The equation of state parameter in terms of radial pressure is also called radial state parameter and is defined as
\begin{equation}
w = \frac{p_r}{\rho}.
\end{equation}
The mass function is defined as
\begin{equation} \label{mass}
m=\int_{r_0}^{r}4\pi r^2\rho dr.
\end{equation}
For radial pressure $p_r$ and tangential pressure $p_t$, the anisotropy parameter is defined as
\begin{equation}
\triangle=p_t-p_r.
\end{equation}

The geometry is attractive or repulsive in nature according as $\triangle$ is negative or positive. If $\triangle = 0$ , then the geometry has an isotropic pressure.

%The metric \ref{metric} is spherical and symmetric.
Embedding diagrams can be used to signify a wormhole and we can get some useful information about the choice of the shape function $b(r)$. For the embedding of a two dimensional surface in a three dimensional Euclidean space, without loss of generality one may consider an equilateral slice $\theta = \frac{\pi}{2}$. For a fixed moment of time i. e. $t$ is constant, the line element \ref{metric}, becomes
\begin{equation} \label{tcons}
ds^2=\frac{1}{1-\frac{b(r)}{r}}dr^2+r^2d\phi^2
\end{equation}
To envisage this, one can embeds this metric into 3-dimensional Euclidean space, whose metric can be written in cartesian and cylindrical coordinates as follows
\begin{equation}
ds^2=dx^2+dy^2+dz^2,
\end{equation}

\begin{equation} \label{cyl}
ds^2=dr^2+r^2d\phi^2+dz^2.
\end{equation}
For an axially symmetric embeded surface $z$ will be a function of radial coordinate $r$ only. Therefore, from Eq. \ref{cyl},
\begin{equation} \label{cyl1}
ds^2=(1+(\frac{dz}{dr})^2)dr^2+r^2d\phi^2.
\end{equation}
From Equations \ref{tcons} and \ref{cyl1}, we can have the equation for embedding surface, defined by
\begin{equation}
\frac{dz}{dr}=\pm\Big(\frac{r}{b(r)}-1\Big)^{-1/2}
\end{equation}
For the existence of a wormhole solution, the geometry follows minimum radius at the throat i. e. $r=b(r)=r_0$ and the embedded surface is vertical i. e.
$\frac{dz}{dr}\to \infty$. The outside space which is far from the mouth of the wormhole is asymptotically flat. For the upper part of the wormhole $z>0$, the proper radial distance can be defined as
\begin{equation}\label{}
  l(r)=\int_{r_0}^{r}\left(\frac{r-b(r)}{r}\right)^{-\frac{1}{2}}dr.
\end{equation}
Similarly, For the lower part of the wormhole $z<0$, the proper radial distance can be defined as
\begin{equation}\label{}
  l(r)=-\int_{r_0}^{r}\left(\frac{r-b(r)}{r}\right)^{-\frac{1}{2}}dr.
\end{equation}
We can also verify that the function $\frac{r-b(r)}{r}$ must be greater than zero for the existence of proper distance. This implies that the shape function $b(r)$ must satisfies $\frac{b(r)}{r}<1$. This fact plays an important role in constructing a specific type of wormhole solutions. The throat flares out condition is necessary for the solution of the wormhole. Mathematically, this flaring out condition needs that the inverse of the embedding function $r(z)$ must satisfy $\frac{d^2r}{dz^2}>0$ near the throat $r_0$.
\begin{figure}
	\centering
\includegraphics[scale=.5]{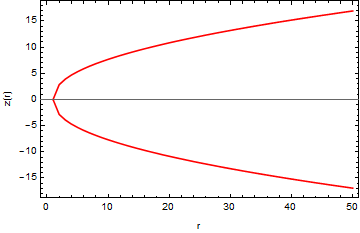}
\end{figure}

The energy conditions play an important role for the existence of a traversable wormhole. The null energy condition is violated over the finite range near the throat of the wormhole. This automatically implies violations of the weak, strong and dominate energy conditions at least in this same range. Thus, the throat of the wormhole must be filled with exotic matter.

%\section{Energy Conditions}
%The important energy conditions are the Null Energy Condition (NEC), Weak Energy Condition (WEC), Strong Energy Condition (SEC) and Dominated Energy Condition (DEC). These conditions are expressed as:
%\begin{itemize}
%	\item [(I)] $\rho + p_r\geq 0$, $\rho + p_t\geq 0$
%	\item [(II)] $\rho \geq 0$, $\rho + p_r> 0$, $\rho + p_t> 0$
%	\item [(III)] $\rho + p_r\geq 0$, $\rho + p_t\geq 0$,  $\rho + p_r +2p_t\geq 0$
%	\item [(IV)] $\rho - \lvert p_r\rvert \geq 0$, $\rho - \lvert p_t\rvert \geq 0$
%\end{itemize}
%%NEC : $\rho + p_r\geq 0$, $\rho + p_t\geq 0$   \\
%%WEC : $\rho \geq 0$, $\rho + p_r> 0$, $\rho + p_t> 0$   \\
%%SEC : $\rho + p_r\geq 0$, $\rho + p_t\geq 0$,  $\rho + p_r +2p_t\geq 0$  \\
%%DEC : $\rho - \lvert p_r\rvert \geq 0$, $\rho - \lvert p_t\rvert \geq 0$\\
%
%A normal matter always satisfies these energy conditions because it possesses positive pressure and positive energy density. The wormholes are non-vacuum solutions of Einstein's field equations and according to Einstein's field theory, they are filled with a matter which is different from the normal matter and is known as exotic matter. This matter does not validate the energy conditions.
%
\section{Wormhole Solutions in $f(R)$ Model}
Nojiri and Odintsov \cite{noj1} introduced an $f(R)$ model, unifying early times inflation and present cosmic acceleration, with the function $f(R)=R-\frac{\alpha}{(R-\lambda_1)^n}+\beta (R-\lambda_2)^m$.
Defining the auxiliary fields $P$ and $Q$, the action can be written as:
\begin{equation}\label{odaction}
S=\frac{1}{\kappa^2}\int \sqrt{-g}[Q(R-P)+f(P)]d^4x.
\end{equation}
Taking the variation with respect to $P$, $Q=f'(P)$, which can be evaluated for $P$ as $P=g(Q)$.
Then action in Eq. \eqref{odaction} takes the form
\begin{equation}\label{action1}
S=\frac{1}{\kappa^2}\int \sqrt{-g}[Q(R-g(Q))+f(g(Q))]d^4x.
\end{equation}
Using $Q=f'(P)$,
\begin{equation}\label{action2}
S=\frac{1}{\kappa^2}\int \sqrt{-g}[f'(P)(R-P)+f(P)].
\end{equation}
Considering conformal transformation
$g^{\mu\nu}\to e^{\sigma}g_{\mu\nu}$, $d$-dimensional scalar curvature is transformed as
\begin{equation}\label{}
R^{(d)}\to e^{-\sigma}\left(R^{(d)}-(d-1)\square\sigma-\frac{(d-1)(d-2)}{4}g^{\mu\nu}\partial_{\mu}\sigma\partial_{\nu}\sigma\right).
\end{equation}
For $d=4$, taking $\sigma=-\ln f'(P)$, the action \eqref{action2} comes out to be
\begin{equation}\label{}
S_{E}=\frac{1}{k^2}\int \sqrt{-g}\bigg[R-\frac{3}{2}\left(\frac{f''(P)}{f'(P)}\right)^2g^{\rho\sigma}\partial_{\rho}P\partial_{\sigma}P
-\frac{Pf'(P)-f(P)}{f'(P)^2}\bigg]d^4x.
\end{equation}
For $\sigma=-\ln (f'(P))=-\ln Q$, the action is
\begin{equation}\label{}
S_{E}=\frac{1}{\kappa^2}\int \sqrt{-g}\bigg[R-\frac{3}{2}g^{\rho\sigma}\partial_{\rho}\sigma\partial_{\sigma}\sigma -Q_1(\sigma)\bigg]d^4x,
\end{equation}
where $Q_1(\sigma)=e^{\sigma}g(e^{-\sigma})-e^{2\sigma} f(g(e^{-\sigma}))=\frac{Pf'(P)-f(P)}{f'(P)^2}$.
As a particular choice of the function $f(R)$,
Nojiri and Odintsov \cite{noj1} considered the form
\begin{equation}\label{}
f(R)=R-\frac{\alpha}{(R-\lambda_1)^n}+\beta (R-\lambda_2)^m,
\end{equation}
where $\alpha, \beta, m$ and $n$ are positive real numbers.

In this work, the $f(R)$ model defined by Nojiri and Odintsov \cite{noj1} is taken in the following form:
\begin{equation}\label{f}
f(R) = R + \alpha R^m - \beta R^{-n},
\end{equation} where $\alpha, \beta, m$ and $n$ are positive constants.
As the curvature increases, the term $R^{-n}$ vanishes and  the term $R^m$ dominates that shows the evolution of inflation at early times.  the value of curvature decreases, the term  $R^m$ vanishes and the term $R^{-n}$ dominate and represents the production of present acceleration. Cao et al. \cite{cao} considered this form of $f(R)$ for the exploration of the probabilities of late time acceleration in the framework of FRW universe. They used state finder diagnostic for models undertaken and found  the best fit models to have evolution from decelerating stage to accelerating stage.   Moradpour \cite{morad} studied traversable wormholes in the background of general relativity and Lyra manifold. He investigated the energy conditions using the shape function
\begin{equation}\label{shape}
b(r)=\frac{r_0 \tanh(r)}{\tanh(r_0)}.
\end{equation}
Here, this shape function is taken into account to obtain the wormhole solutions.
The energy density, radial and tangential pressures obtained from Eqs. \ref{6}-\ref{8} are as follows:

\begin{eqnarray}\label{d}
\rho&=&\frac{1}{4 r^2 {r_0}}\Bigg[-2^{2-n} \cosh^2(r) (2 r \tanh (r)+1) \tanh ({r_0}) (r-{r_0} \tanh (r) \coth ({r_0}))\nonumber\\
&\times& \left(\frac{{r_0} {\sech}^2(r) \coth ({r_0})}{r}\right)^{-n} \left(\alpha  (m-1) m \left(-2^{m+n}\right) \left(\frac{{r_0} {\sech}^2(r) \coth ({r_0})}{r}\right)^{m+n}+\beta  n^2\right.\nonumber\\
&+&\left.\beta  n\right)-2^{-n} {r_0} (2 r \tanh (r)+1) (r-\sinh (r) \cosh (r)) \left(\frac{{r_0} {\sech}^2(r) \coth ({r_0})}{r}\right)^{-n} \nonumber\\
&\times&\left(\alpha  (m-1) m \left(-2^{m+n}\right) \left(\frac{{r_0} {\sech}^2(r) \coth ({r_0})}{r}\right)^{m+n}+\beta  n^2+ \beta  n\right)-2 r {r_0} \left(\alpha  2^m m r\right.\nonumber\\
 &\times&\left.\left(\frac{{r_0} {\sech}^2(r) \coth ({r_0})}{r}\right)^m+\beta  2^{-n} n r \left(\frac{{r_0} {\sech}^2(r) \coth ({r_0})}{r}\right)^{-n}+2 {r_0} {\sech}^2(r) \coth ({r_0})\right)\nonumber\\
 &+&2 {r_0} \left(\alpha  2^m m r \left(\frac{{r_0} {\sech}^2(r) \coth ({r_0})}{r}\right)^m+\beta  2^{-n} n r \left(\frac{{r_0} {\sech}^2(r) \coth ({r_0})}{r}\right)^{-n}+2 {r_0} {\sech}^2(r) \right.\nonumber\\
&\times& \left.\coth ({r_0})\right) +2 r {r_0} \left(\alpha  2^m r \left(\frac{{r_0} {\sech}^2(r) \coth ({r_0})}{r}\right)^m-\beta  2^{-n} r \left(\frac{{r_0} {\sech}^2(r) \coth ({r_0})}{r}\right)^{-n}\right.\nonumber\\
&+&\left.2 {r_0} {\sech}^2(r) \coth ({r_0})\right)+\alpha  2^{m+2} (1-m) m \tanh ({r_0}) \left(\cosh ^2(r)+r (-2 r+\sinh (2 r)+r \cosh (2 r))\right)\nonumber\\
&\times& (r-{r_0} \tanh (r) \coth ({r_0})) \left(\frac{{r_0} {\sech}^2(r) \coth ({r_0})}{r}\right)^m+\alpha  2^{m+1} (m-2) (m-1) m \tanh ({r_0}) (2 r \sinh (r)\nonumber\\
&+&\cosh (r))^2 ({r_0} \tanh (r) \coth ({r_0})-r) \left(\frac{{r_0} {\sech}^2(r) \coth ({r_0})}{r}\right)^m+\beta  2^{2-n} n (n+1) \tanh ({r_0}) \left(\cosh ^2(r)\right.\nonumber
%&+&\left.r (-2 r+\sinh (2 r)+r \cosh (2 r))\right) (r-{r_0} \tanh (r) \coth ({r_0})) \left(\frac{{r_0} {\sech}^2(r) \coth ({r_0})}{r}\right)^{-n}\nonumber\\
%&+&\beta  2^{1-n} n (n+1) (n+2) \tanh ({r_0}) (2 r \sinh (r)+\cosh (r))^2 ({r_0} \tanh (r) \coth ({r_0})-r) \nonumber\\
%&\times&\left(\frac{{r_0} {\sech}^2(r) \coth ({r_0})}{r}\right)^{-n}\Bigg]\nonumber
\end{eqnarray}
\begin{eqnarray}
&+&\cosh (r))^2 ({r_0} \tanh (r) \coth ({r_0})-r) \left(\frac{{r_0} {\sech}^2(r) \coth ({r_0})}{r}\right)^m+\beta  2^{2-n} n (n+1) \tanh ({r_0}) \left(\cosh ^2(r)\right.\nonumber\\
&+&\left.r (-2 r+\sinh (2 r)+r \cosh (2 r))\right) (r-{r_0} \tanh (r) \coth ({r_0})) \left(\frac{{r_0} {\sech}^2(r) \coth ({r_0})}{r}\right)^{-n}\nonumber\\
&+&\beta  2^{1-n} n (n+1) (n+2) \tanh ({r_0}) (2 r \sinh (r)+\cosh (r))^2 ({r_0} \tanh (r) \coth ({r_0})-r) \nonumber\\
&\times&\left(\frac{{r_0} {\sech}^2(r) \coth ({r_0})}{r}\right)^{-n}\Bigg]
\end{eqnarray}

\begin{eqnarray}\label{pr}
p_r&=&\frac{1}{r^3 {r_0}}\Bigg[2^{-n-1} \left(\frac{{r_0} {\sech}^2(r) \coth ({r_0})}{r}\right)^{-n} \left({r_0} \left(\alpha  m^2 r 2^{m+n} \sinh (2 r) \left(\frac{{r_0} {\sech}^2(r) \coth ({r_0})}{r}\right)^{m+n}\right.\right.\nonumber\\
&-&\left.\left.4 r^2 \sinh ^2(r) \left(\alpha  (m-1) m \left(-2^{m+n}\right) \left(\frac{{r_0} {\sech}^2(r) \coth ({r_0})}{r}\right)^{m+n}+\beta  n^2+\beta  n\right)-\alpha  r^3 2^{m+n}\right.\right. \nonumber\\
&\times&\left(\frac{{r_0} {\sech}^2(r) \coth ({r_0})}{r}\right)^{m+n}+\alpha  m r^3 2^{m+n} \left(\frac{{r_0} {\sech}^2(r) \coth ({r_0})}{r}\right)^{m+n}+\beta  n r^3-2^{n+1} {r_0}\nonumber\\
&\times& \left.\left.\tanh (r) \coth ({r_0}) \left(\frac{{r_0} {\sech}^2(r) \coth ({r_0})}{r}\right)^n+\beta  r^3\right)+2 r^2 \cosh ^2(r) \tanh ({r_0}) \left(\alpha  (m-1) m \right.\right.\nonumber\\
&\times& \left.\left.\left(-2^{m+n}\right) \left(\frac{{r_0} {\sech}^2(r) \coth ({r_0})}{r}\right)^{m+n}+\beta  n^2+\beta  n\right)+r \sinh (r) \cosh (r) \tanh ({r_0}) \left(4 r^2 \right.\right.\nonumber\\
&\times&\left.\left.\left(\alpha  (m-1) m \left(-2^{m+n}\right) \left(\frac{{r_0} {\sech}^2(r) \coth ({r_0})}{r}\right)^{m+n}+\beta  n^2+\beta  n\right)-{r_0} \coth ({r_0}) \left(3 \alpha  m 2^{m+n}\right.\right.\right. \nonumber\\
&\times&\left.\left.\left.\left(\frac{{r_0} {\sech}^2(r) \coth ({r_0})}{r}\right)^{m+n}+2 \beta  n^2+3 \beta  n\right)\right)\right)\Bigg]
\end{eqnarray}

\begin{eqnarray}\label{pt}
p_t&=&\frac{1}{4 r^3 {r_0}}\Bigg[2^{-n} r \tanh ({r_0}) \left(\frac{{r_0} {\sech}^2(r) \coth ({r_0})}{r}\right)^{-n} \left(\sinh (r) \cosh (r) \left(4 r^2 \left(\alpha  m \left(2 m^2-5 m\right.\right.\right.\right.\nonumber\\
&+&\left.\left.\left.\left.3\right) 2^{m+n} \left(\frac{{r_0} {\sech}^2(r) \coth ({r_0})}{r}\right)^{m+n}+2 \beta  n^3+5 \beta  n^2\right)-{r_0} \coth ({r_0}) \left(-5 \alpha  m^2 2^{m+n} \right.\right.\right.\nonumber\\
&+&\left.\left.\left.\left(\frac{{r_0} {\sech}^2(r) \coth ({r_0})}{r}\right)^{m+n}+2 \beta  n^3+5 \beta  n^2+2 \beta  n\right)\right)-{r_0} \coth ({r_0}) \left(r \left(\alpha  m^2 2^{m+n}\right.\right.\right.\nonumber\\
&\times& \left.\left.\left.\left(\frac{{r_0} {\sech}^2(r) \coth ({r_0})}{r}\right)^{m+n}+2 r \tanh (r) \left(\alpha  (m-1) m \left(-2^{m+n}\right) \left(\frac{{r_0} {\sech}^2(r) \coth ({r_0})}{r}\right)^{m+n}\right.\right.\right.\right.\nonumber\\
&+&\left.\left.\left.\left.\beta  n^2+\beta  n\right)+\alpha  r 2^{m+n+1} \left(\frac{{r_0} {\sech}^2(r) \coth ({r_0})}{r}\right)^{m+n}-\alpha  m r 2^{m+n+1} \left(\frac{{r_0} {\sech}^2(r) \coth ({r_0})}{r}\right)^{m+n}\right.\right.\right.\nonumber
\end{eqnarray}

\begin{eqnarray}
&-&\left.\left.\left.\beta  n^2-2 \beta  n r-2 \beta  r\right)+2 r \sinh ^2(r) \left(\alpha  m \left(4 m^2-11 m+7\right) 2^{m+n} \left(\frac{{r_0} {\sech}^2(r) \coth ({r_0})}{r}\right)^{m+n}\right.\right.\right.\nonumber\\
&+&\left.\left.\left.4 r \tanh (r) \left(\alpha  (m-1)^2 m 2^{m+n} \left(\frac{{r_0} {\sech}^2(r) \coth ({r_0})}{r}\right)^{m+n}+\beta  n^3+2 \beta  n^2+\beta  n\right)+4 \beta  n^3\right.\right.\right.\nonumber\\
&+&\left.\left.\left.11 \beta  n^2+7 \beta  n\right)+\alpha  m \left(m^2+1\right) 2^{m+n} \sinh (2 r) \left(\frac{{r_0} {\sech}^2(r) \coth ({r_0})}{r}\right)^{m+n}\right)+4 r^2 \left(r\right.\right.
\nonumber\\
&\times& \left.\left.
\left(2 \sinh ^2(r) \left(\alpha  (m-1)^2 m 2^{m+n} \left(\frac{{r_0} {sech}^2(r) \coth ({r_0})}{r}\right)^{m+n}+\beta  n^3+2 \beta  n^2+\beta  n\right)+\alpha  (m-1) m \right.\right.\right.\nonumber\\
&\times&  \left.\left.\left.\left(-2^{m+n}\right) \left(\frac{{r_0} {\sech}^2(r) \coth ({r_0})}{r}\right)^{m+n}+\beta  n^2+\beta  n\right)+3 \beta  n \sinh (r) \cosh (r)\right)+2 r \cosh ^2(r) \right.\nonumber\\
&\times&\left.\left(\alpha  (m-1)^2 m 2^{m+n} \left(\frac{{r_0} {\sech}^2(r) \coth ({r_0})}{r}\right)^{m+n}+\beta  n^3+2 \beta  n^2+\beta  n\right)\right)\nonumber\\
&+&{r_0}^2 (\sinh (2 r)-2 r) {\sech}^2(r) \coth ({r_0})\Bigg]
\end{eqnarray}
Using Equations \ref{d}-\ref{pt}, the null energy condition terms $\rho+p_r$ \& $\rho + p_t$ and the dominated energy condition terms $\rho-|p_r|$ \& $\rho-|p_t|$ are obtained as:

\begin{eqnarray}\label{}
	\rho+p_r&=&\frac{1}{4 r^2 {r_0}}\Bigg[-2^{2-n} \cosh^2(r) (2 r \tanh (r)+1) \tanh ({r_0}) (r-{r_0} \tanh (r) \coth ({r_0}))\nonumber\\
	&\times& \left(\frac{{r_0} {\sech}^2(r) \coth ({r_0})}{r}\right)^{-n} \left(\alpha  (m-1) m \left(-2^{m+n}\right) \left(\frac{{r_0} {\sech}^2(r) \coth ({r_0})}{r}\right)^{m+n}+\beta  n^2\right.\nonumber\\
	&+&\left.\beta  n\right)-2^{-n} {r_0} (2 r \tanh (r)+1) (r-\sinh (r) \cosh (r)) \left(\frac{{r_0} {\sech}^2(r) \coth ({r_0})}{r}\right)^{-n} \nonumber\\
	&\times&\left(\alpha  (m-1) m \left(-2^{m+n}\right) \left(\frac{{r_0} {\sech}^2(r) \coth ({r_0})}{r}\right)^{m+n}+\beta  n^2+ \beta  n\right)-2 r {r_0} \left(\alpha  2^m m r\right.\nonumber\\
	&\times&\left.\left(\frac{{r_0} {\sech}^2(r) \coth ({r_0})}{r}\right)^m+\beta  2^{-n} n r \left(\frac{{r_0} {\sech}^2(r) \coth ({r_0})}{r}\right)^{-n}+2 {r_0} {\sech}^2(r) \coth ({r_0})\right)\nonumber\\
	&+&2 {r_0} \left(\alpha  2^m m r \left(\frac{{r_0} {\sech}^2(r) \coth ({r_0})}{r}\right)^m+\beta  2^{-n} n r \left(\frac{{r_0} {\sech}^2(r) \coth ({r_0})}{r}\right)^{-n}+2 {r_0} {\sech}^2(r) \right.\nonumber\\
	&\times& \left.\coth ({r_0})\right) +2 r {r_0} \left(\alpha  2^m r \left(\frac{{r_0} {\sech}^2(r) \coth ({r_0})}{r}\right)^m-\beta  2^{-n} r \left(\frac{{r_0} {\sech}^2(r) \coth ({r_0})}{r}\right)^{-n}\right.\nonumber\\
	&+&\left.2 {r_0} {\sech}^2(r) \coth ({r_0})\right)+\alpha  2^{m+2} (1-m) m \tanh ({r_0}) \left(\cosh ^2(r)+r (-2 r+\sinh (2 r)\right.\nonumber
		\end{eqnarray}
	\begin{eqnarray}
	&+&\left.+r \cosh (2 r))\right) (r-{r_0} \tanh (r) \coth ({r_0})) \left(\frac{{r_0} {\sech}^2(r) \coth ({r_0})}{r}\right)^m+\alpha  2^{m+1} (m-2) \nonumber\\
		&\times&(m-1) m \tanh ({r_0}) (2 r \sinh (r)+\cosh (r))^2 ({r_0} \tanh (r) \coth ({r_0})-r) \left(\frac{{r_0} {\sech}^2(r) \coth ({r_0})}{r}\right)^m\nonumber\\
	&+&\beta  2^{2-n} n (n+1) \tanh ({r_0}) \left(\cosh ^2(r)\right.+\left.r (-2 r+\sinh (2 r)+r \cosh (2 r))\right) (r-{r_0} \tanh (r)\nonumber\\
&\times& \coth ({r_0})) \left(\frac{{r_0} {\sech}^2(r) \coth ({r_0})}{r}\right)^{-n}+\beta  2^{1-n} n (n+1) (n+2) \tanh ({r_0}) (2 r \sinh (r) \nonumber\\
&+&\cosh (r))^2 ({r_0} \tanh (r) \coth ({r_0})-r)\left(\frac{{r_0} {\sech}^2(r) \coth ({r_0})}{r}\right)^{-n}\Bigg]\nonumber\\
&+&\frac{1}{r^3 {r_0}}\Bigg[2^{-n-1} \left(\frac{{r_0} {\sech}^2(r) \coth ({r_0})}{r}\right)^{-n} \left({r_0} \left(\alpha  m^2 r 2^{m+n} \sinh (2 r) \left(\frac{{r_0} {\sech}^2(r) \coth ({r_0})}{r}\right)^{m+n}\right.\right.\nonumber\\
&-&\left.\left.4 r^2 \sinh ^2(r) \left(\alpha  (m-1) m \left(-2^{m+n}\right) \left(\frac{{r_0} {sech}^2(r) \coth ({r_0})}{r}\right)^{m+n}+\beta  n^2+\beta  n\right)-\alpha  r^3 2^{m+n}\right.\right. \nonumber\\
&\times&\left(\frac{{r_0} {\sech}^2(r) \coth ({r_0})}{r}\right)^{m+n}+\alpha  m r^3 2^{m+n} \left(\frac{{r_0} {\sech}^2(r) \coth ({r_0})}{r}\right)^{m+n}+\beta  n r^3-2^{n+1} {r_0}\nonumber\\
&\times& \left.\left.\tanh (r) \coth ({r_0}) \left(\frac{{r_0} {\sech}^2(r) \coth ({r_0})}{r}\right)^n+\beta  r^3\right)+2 r^2 \cosh ^2(r) \tanh ({r_0}) \left(\alpha  (m-1) m \right.\right.\nonumber\\
&\times& \left.\left.\left(-2^{m+n}\right) \left(\frac{{r_0} {\sech}^2(r) \coth ({r_0})}{r}\right)^{m+n}+\beta  n^2+\beta  n\right)+r \sinh (r) \cosh (r) \tanh ({r_0}) \left(4 r^2 \right.\right.\nonumber\\
&\times&\left.\left.\left(\alpha  (m-1) m \left(-2^{m+n}\right) \left(\frac{{r_0} {\sech}^2(r) \coth ({r_0})}{r}\right)^{m+n}+\beta  n^2+\beta  n\right)-{r_0} \coth ({r_0}) \left(3 \alpha  m 2^{m+n}\right.\right.\right. \nonumber\\
&\times&\left.\left.\left.\left(\frac{{r_0} {\sech}^2(r) \coth ({r_0})}{r}\right)^{m+n}+2 \beta  n^2+3 \beta  n\right)\right)\right)\Bigg]
\end{eqnarray}

\begin{eqnarray}\label{}
\rho+p_t&=&\frac{1}{4 r^2 {r_0}}\Bigg[-2^{2-n} \cosh^2(r) (2 r \tanh (r)+1) \tanh ({r_0}) (r-{r_0} \tanh (r) \coth ({r_0}))\nonumber\\
&\times& \left(\frac{{r_0} {\sech}^2(r) \coth ({r_0})}{r}\right)^{-n} \left(\alpha  (m-1) m \left(-2^{m+n}\right) \left(\frac{{r_0} {\sech}^2(r) \coth ({r_0})}{r}\right)^{m+n}+\beta  n^2\right.\nonumber\\
&+&\left.\beta  n\right)-2^{-n} {r_0} (2 r \tanh (r)+1) (r-\sinh (r) \cosh (r)) \left(\frac{{r_0} {\sech}^2(r) \coth ({r_0})}{r}\right)^{-n} \nonumber\\
&\times&\left(\alpha  (m-1) m \left(-2^{m+n}\right) \left(\frac{{r_0} {\sech}^2(r) \coth ({r_0})}{r}\right)^{m+n}+\beta  n^2+ \beta  n\right)-2 r {r_0} \left(\alpha  2^m m r\right.\nonumber\\
&\times&\left.\left(\frac{{r_0} {\sech}^2(r) \coth ({r_0})}{r}\right)^m+\beta  2^{-n} n r \left(\frac{{r_0} {\sech}^2(r) \coth ({r_0})}{r}\right)^{-n}+2 {r_0} {\sech}^2(r) \coth ({r_0})\right)\nonumber
\end{eqnarray}

\begin{eqnarray}
&+&2 {r_0} \left(\alpha  2^m m r \left(\frac{{r_0} {\sech}^2(r) \coth ({r_0})}{r}\right)^m+\beta  2^{-n} n r \left(\frac{{r_0} {\sech}^2(r) \coth ({r_0})}{r}\right)^{-n}+2 {r_0} {\sech}^2(r) \right.\nonumber\\
&\times& \left.\coth ({r_0})\right) +2 r {r_0} \left(\alpha  2^m r \left(\frac{{r_0} {\sech}^2(r) \coth ({r_0})}{r}\right)^m-\beta  2^{-n} r \left(\frac{{r_0} {\sech}^2(r) \coth ({r_0})}{r}\right)^{-n}\right.\nonumber\\
&+&\left.2 {r_0} {\sech}^2(r) \coth ({r_0})\right)+\alpha  2^{m+2} (1-m) m \tanh ({r_0}) \left(\cosh ^2(r)+r (-2 r+\sinh (2 r)+r \cosh (2 r))\right)\nonumber\\
&\times& (r-{r_0} \tanh (r) \coth ({r_0})) \left(\frac{{r_0} {\sech}^2(r) \coth ({r_0})}{r}\right)^m+\alpha  2^{m+1} (m-2) (m-1) m \tanh ({r_0}) (2 r \sinh (r)\nonumber\\
&+&\cosh (r))^2 ({r_0} \tanh (r) \coth ({r_0})-r) \left(\frac{{r_0} {\sech}^2(r) \coth ({r_0})}{r}\right)^m+\beta  2^{2-n} n (n+1) \tanh ({r_0}) \left(\cosh ^2(r)\right.\nonumber\\
&+&\left.r (-2 r+\sinh (2 r)+r \cosh (2 r))\right) (r-{r_0} \tanh (r) \coth ({r_0})) \left(\frac{{r_0} {\sech}^2(r) \coth ({r_0})}{r}\right)^{-n}\nonumber\\
&+&\beta  2^{1-n} n (n+1) (n+2) \tanh ({r_0}) (2 r \sinh (r)+\cosh (r))^2 ({r_0} \tanh (r) \coth ({r_0})-r) \nonumber\\
&\times&\left(\frac{{r_0} {\sech}^2(r) \coth ({r_0})}{r}\right)^{-n}\Bigg]\nonumber\\
&+&\frac{1}{4 r^3 {r_0}}\Bigg[2^{-n} r \tanh ({r_0}) \left(\frac{{r_0} {\sech}^2(r) \coth ({r_0})}{r}\right)^{-n} \left(\sinh (r) \cosh (r) \left(4 r^2 \left(\alpha  m \left(2 m^2-5 m\right.\right.\right.\right.\nonumber\\
&+&\left.\left.\left.\left.3\right) 2^{m+n} \left(\frac{{r_0} {\sech}^2(r) \coth ({r_0})}{r}\right)^{m+n}+2 \beta  n^3+5 \beta  n^2\right)-{r_0} \coth ({r_0}) \left(-5 \alpha  m^2 2^{m+n} \right.\right.\right.\nonumber\\
&+&\left.\left.\left.\left(\frac{{r_0} {\sech}^2(r) \coth ({r_0})}{r}\right)^{m+n}+2 \beta  n^3+5 \beta  n^2+2 \beta  n\right)\right)-{r_0} \coth ({r_0}) \left(r \left(\alpha  m^2 2^{m+n}\right.\right.\right.\nonumber\\
&\times& \left.\left.\left.\left(\frac{{r_0} {\sech}^2(r) \coth ({r_0})}{r}\right)^{m+n}+2 r \tanh (r) \left(\alpha  (m-1) m \left(-2^{m+n}\right) \left(\frac{{r_0} {\sech}^2(r) \coth ({r_0})}{r}\right)^{m+n}\right.\right.\right.\right.\nonumber\\
&+&\left.\left.\left.\left.\beta  n^2+\beta  n\right)+\alpha  r 2^{m+n+1} \left(\frac{{r_0} {\sech}^2(r) \coth ({r_0})}{r}\right)^{m+n}-\alpha  m r 2^{m+n+1} \left(\frac{{r_0} {\sech}^2(r) \coth ({r_0})}{r}\right)^{m+n}\right.\right.\right.\nonumber\\
&-&\left.\left.\left.\beta  n^2-2 \beta  n r-2 \beta  r\right)+2 r \sinh ^2(r) \left(\alpha  m \left(4 m^2-11 m+7\right) 2^{m+n} \left(\frac{{r_0} {\sech}^2(r) \coth ({r_0})}{r}\right)^{m+n}\right.\right.\right.\nonumber\\
&+&\left.\left.\left.4 r \tanh (r) \left(\alpha  (m-1)^2 m 2^{m+n} \left(\frac{{r_0} {\sech}^2(r) \coth ({r_0})}{r}\right)^{m+n}+\beta  n^3+2 \beta  n^2+\beta  n\right)+4 \beta  n^3\right.\right.\right.\nonumber\\
&+&\left.\left.\left.11 \beta  n^2+7 \beta  n\right)+\alpha  m \left(m^2+1\right) 2^{m+n} \sinh (2 r) \left(\frac{{r_0} {\sech}^2(r) \coth ({r_0})}{r}\right)^{m+n}\right)+4 r^2 \left(r\right.\right.
\nonumber\\
&\times& \left.\left.
 \left(2 \sinh ^2(r) \left(\alpha  (m-1)^2 m 2^{m+n} \left(\frac{{r_0} {\sech}^2(r) \coth ({r_0})}{r}\right)^{m+n}+\beta  n^3+2 \beta  n^2+\beta  n\right)+\alpha  (m-1) m \right.\right.\right.\nonumber\\
&\times&  \left.\left.\left.\left(-2^{m+n}\right) \left(\frac{{r_0} {\sech}^2(r) \coth ({r_0})}{r}\right)^{m+n}+\beta  n^2+\beta  n\right)+3 \beta  n \sinh (r) \cosh (r)\right)+2 r \cosh ^2(r)\right. \nonumber
\end{eqnarray}
\begin{eqnarray}
&\times&\left.\left(\alpha  (m-1)^2 m 2^{m+n} \left(\frac{{r_0} {\sech}^2(r) \coth ({r_0})}{r}\right)^{m+n}+\beta  n^3+2 \beta  n^2+\beta  n\right)\right)+{r_0}^2 (\sinh (2 r)\nonumber\\
&-&2 r) {\sech}^2(r) \coth ({r_0})\Bigg]
\end{eqnarray}

\begin{eqnarray}\label{}
\rho-|p_r|&=&\frac{1}{4 r^2 {r_0}}\Bigg[-2^{2-n} \cosh^2(r) (2 r \tanh (r)+1) \tanh ({r_0}) (r-{r_0} \tanh (r) \coth ({r_0}))\nonumber\\
&\times& \left(\frac{{r_0} {\sech}^2(r) \coth ({r_0})}{r}\right)^{-n} \left(\alpha  (m-1) m \left(-2^{m+n}\right) \left(\frac{{r_0} {\sech}^2(r) \coth ({r_0})}{r}\right)^{m+n}+\beta  n^2\right.\nonumber\\
&+&\left.\beta  n\right)-2^{-n} {r_0} (2 r \tanh (r)+1) (r-\sinh (r) \cosh (r)) \left(\frac{{r_0} {\sech}^2(r) \coth ({r_0})}{r}\right)^{-n} \nonumber\\
&\times&\left(\alpha  (m-1) m \left(-2^{m+n}\right) \left(\frac{{r_0} {\sech}^2(r) \coth ({r_0})}{r}\right)^{m+n}+\beta  n^2+ \beta  n\right)-2 r {r_0} \left(\alpha  2^m m r\right.\nonumber\\
&\times&\left.\left(\frac{{r_0} {\sech}^2(r) \coth ({r_0})}{r}\right)^m+\beta  2^{-n} n r \left(\frac{{r_0} {\sech}^2(r) \coth ({r_0})}{r}\right)^{-n}+2 {r_0} {\sech}^2(r) \coth ({r_0})\right)\nonumber\\
&+&2 {r_0} \left(\alpha  2^m m r \left(\frac{{r_0} {\sech}^2(r) \coth ({r_0})}{r}\right)^m+\beta  2^{-n} n r \left(\frac{{r_0} {\sech}^2(r) \coth ({r_0})}{r}\right)^{-n}+2 {r_0} {\sech}^2(r) \right.\nonumber\\
&\times& \left.\coth ({r_0})\right) +2 r {r_0} \left(\alpha  2^m r \left(\frac{{r_0} {\sech}^2(r) \coth ({r_0})}{r}\right)^m-\beta  2^{-n} r \left(\frac{{r_0} {\sech}^2(r) \coth ({r_0})}{r}\right)^{-n}\right.\nonumber\\
&+&\left.2 {r_0} {\sech}^2(r) \coth ({r_0})\right)+\alpha  2^{m+2} (1-m) m \tanh ({r_0}) \left(\cosh ^2(r)+r (-2 r+\sinh (2 r)\right.\nonumber\\
&+&\left.+r \cosh (2 r))\right) (r-{r_0} \tanh (r) \coth ({r_0})) \left(\frac{{r_0} {\sech}^2(r) \coth ({r_0})}{r}\right)^m+\alpha  2^{m+1} (m-2) \nonumber\\
&\times&(m-1) m \tanh ({r_0}) (2 r \sinh (r)+\cosh (r))^2 ({r_0} \tanh (r) \coth ({r_0})-r) \left(\frac{{r_0} {\sech}^2(r) \coth ({r_0})}{r}\right)^m\nonumber\\
&+&\beta  2^{2-n} n (n+1) \tanh ({r_0}) \left(\cosh ^2(r)\right.+\left.r (-2 r+\sinh (2 r)+r \cosh (2 r))\right) (r-{r_0} \tanh (r)\nonumber\\
&\times& \coth ({r_0})) \left(\frac{{r_0} {\sech}^2(r) \coth ({r_0})}{r}\right)^{-n}+\beta  2^{1-n} n (n+1) (n+2) \tanh ({r_0}) (2 r \sinh (r) \nonumber\\
&+&\cosh (r))^2 ({r_0} \tanh (r) \coth ({r_0})-r)\left(\frac{{r_0} {\sech}^2(r) \coth ({r_0})}{r}\right)^{-n}\Bigg]\nonumber\\
&-&\Bigg|\frac{1}{r^3 {r_0}}\Bigg[2^{-n-1} \left(\frac{{r_0} {\sech}^2(r) \coth ({r_0})}{r}\right)^{-n} \left({r_0} \left(\alpha  m^2 r 2^{m+n} \sinh (2 r) \left(\frac{{r_0} {\sech}^2(r) \coth ({r_0})}{r}\right)^{m+n}\right.\right.\nonumber\\
&-&\left.\left.4 r^2 \sinh ^2(r) \left(\alpha  (m-1) m \left(-2^{m+n}\right) \left(\frac{{r_0} {sech}^2(r) \coth ({r_0})}{r}\right)^{m+n}+\beta  n^2+\beta  n\right)-\alpha  r^3 2^{m+n}\right.\right. \nonumber
\end{eqnarray}
\begin{eqnarray}
&\times&\left(\frac{{r_0} {\sech}^2(r) \coth ({r_0})}{r}\right)^{m+n}+\alpha  m r^3 2^{m+n} \left(\frac{{r_0} {\sech}^2(r) \coth ({r_0})}{r}\right)^{m+n}+\beta  n r^3-2^{n+1} {r_0}\nonumber\\
&\times& \left.\left.\tanh (r) \coth ({r_0}) \left(\frac{{r_0} {\sech}^2(r) \coth ({r_0})}{r}\right)^n+\beta  r^3\right)+2 r^2 \cosh ^2(r) \tanh ({r_0}) \left(\alpha  (m-1) m \right.\right.\nonumber\\
&\times& \left.\left.\left(-2^{m+n}\right) \left(\frac{{r_0} {\sech}^2(r) \coth ({r_0})}{r}\right)^{m+n}+\beta  n^2+\beta  n\right)+r \sinh (r) \cosh (r) \tanh ({r_0}) \left(4 r^2 \right.\right.\nonumber\\
&\times&\left.\left.\left(\alpha  (m-1) m \left(-2^{m+n}\right) \left(\frac{{r_0} {\sech}^2(r) \coth ({r_0})}{r}\right)^{m+n}+\beta  n^2+\beta  n\right)-{r_0} \coth ({r_0}) \left(3 \alpha  m 2^{m+n}\right.\right.\right. \nonumber\\
&\times&\left.\left.\left.\left(\frac{{r_0} {\sech}^2(r) \coth ({r_0})}{r}\right)^{m+n}+2 \beta  n^2+3 \beta  n\right)\right)\right)\Bigg]\Bigg|
\end{eqnarray}

\begin{eqnarray}\label{}
\rho-|p_t|&=&\frac{1}{4 r^2 {r_0}}\Bigg[-2^{2-n} \cosh^2(r) (2 r \tanh (r)+1) \tanh ({r_0}) (r-{r_0} \tanh (r) \coth ({r_0}))\nonumber\\
&\times& \left(\frac{{r_0} {\sech}^2(r) \coth ({r_0})}{r}\right)^{-n} \left(\alpha  (m-1) m \left(-2^{m+n}\right) \left(\frac{{r_0} {\sech}^2(r) \coth ({r_0})}{r}\right)^{m+n}+\beta  n^2\right.\nonumber\\
&+&\left.\beta  n\right)-2^{-n} {r_0} (2 r \tanh (r)+1) (r-\sinh (r) \cosh (r)) \left(\frac{{r_0} {\sech}^2(r) \coth ({r_0})}{r}\right)^{-n} \nonumber\\
&\times&\left(\alpha  (m-1) m \left(-2^{m+n}\right) \left(\frac{{r_0} {\sech}^2(r) \coth ({r_0})}{r}\right)^{m+n}+\beta  n^2+ \beta  n\right)-2 r {r_0} \left(\alpha  2^m m r\right.\nonumber\\
&\times&\left.\left(\frac{{r_0} {\sech}^2(r) \coth ({r_0})}{r}\right)^m+\beta  2^{-n} n r \left(\frac{{r_0} {\sech}^2(r) \coth ({r_0})}{r}\right)^{-n}+2 {r_0} {\sech}^2(r) \coth ({r_0})\right)\nonumber\\
&+&2 {r_0} \left(\alpha  2^m m r \left(\frac{{r_0} {\sech}^2(r) \coth ({r_0})}{r}\right)^m+\beta  2^{-n} n r \left(\frac{{r_0} {\sech}^2(r) \coth ({r_0})}{r}\right)^{-n}+2 {r_0} {\sech}^2(r) \right.\nonumber\\
&\times& \left.\coth ({r_0})\right) +2 r {r_0} \left(\alpha  2^m r \left(\frac{{r_0} {\sech}^2(r) \coth ({r_0})}{r}\right)^m-\beta  2^{-n} r \left(\frac{{r_0} {\sech}^2(r) \coth ({r_0})}{r}\right)^{-n}\right.\nonumber\\
&+&\left.2 {r_0} {\sech}^2(r) \coth ({r_0})\right)+\alpha  2^{m+2} (1-m) m \tanh ({r_0}) \left(\cosh ^2(r)+r (-2 r+\sinh (2 r)+r \cosh (2 r))\right)\nonumber\\
&\times& (r-{r_0} \tanh (r) \coth ({r_0})) \left(\frac{{r_0} {\sech}^2(r) \coth ({r_0})}{r}\right)^m+\alpha  2^{m+1} (m-2) (m-1) m \tanh ({r_0}) (2 r \sinh (r)\nonumber\\
&+&\cosh (r))^2 ({r_0} \tanh (r) \coth ({r_0})-r) \left(\frac{{r_0} {\sech}^2(r) \coth ({r_0})}{r}\right)^m+\beta  2^{2-n} n (n+1) \tanh ({r_0}) \left(\cosh ^2(r)\right.\nonumber\\
&+&\left.r (-2 r+\sinh (2 r)+r \cosh (2 r))\right) (r-{r_0} \tanh (r) \coth ({r_0})) \left(\frac{{r_0} {\sech}^2(r) \coth ({r_0})}{r}\right)^{-n}\nonumber
\end{eqnarray}

\begin{eqnarray}
&+&\beta  2^{1-n} n (n+1) (n+2) \tanh ({r_0}) (2 r \sinh (r)+\cosh (r))^2 ({r_0} \tanh (r) \coth ({r_0})-r) \nonumber\\
&\times&\left(\frac{{r_0} {\sech}^2(r) \coth ({r_0})}{r}\right)^{-n}\Bigg]\nonumber\\
&-&\Bigg|\frac{1}{4 r^3 {r_0}}\Bigg[2^{-n} r \tanh ({r_0}) \left(\frac{{r_0} {\sech}^2(r) \coth ({r_0})}{r}\right)^{-n} \left(\sinh (r) \cosh (r) \left(4 r^2 \left(\alpha  m \left(2 m^2-5 m\right.\right.\right.\right.\nonumber\\
&+&\left.\left.\left.\left.3\right) 2^{m+n} \left(\frac{{r_0} {\sech}^2(r) \coth ({r_0})}{r}\right)^{m+n}+2 \beta  n^3+5 \beta  n^2\right)-{r_0} \coth ({r_0}) \left(-5 \alpha  m^2 2^{m+n} \right.\right.\right.\nonumber\\
&+&\left.\left.\left.\left(\frac{{r_0} {\sech}^2(r) \coth ({r_0})}{r}\right)^{m+n}+2 \beta  n^3+5 \beta  n^2+2 \beta  n\right)\right)-{r_0} \coth ({r_0}) \left(r \left(\alpha  m^2 2^{m+n}\right.\right.\right.\nonumber\\
&\times& \left.\left.\left.\left(\frac{{r_0} {\sech}^2(r) \coth ({r_0})}{r}\right)^{m+n}+2 r \tanh (r) \left(\alpha  (m-1) m \left(-2^{m+n}\right) \left(\frac{{r_0} {\sech}^2(r) \coth ({r_0})}{r}\right)^{m+n}\right.\right.\right.\right.\nonumber\\
&+&\left.\left.\left.\left.\beta  n^2+\beta  n\right)+\alpha  r 2^{m+n+1} \left(\frac{{r_0} {\sech}^2(r) \coth ({r_0})}{r}\right)^{m+n}-\alpha  m r 2^{m+n+1} \left(\frac{{r_0} {\sech}^2(r) \coth ({r_0})}{r}\right)^{m+n}\right.\right.\right.\nonumber\\
&-&\left.\left.\left.\beta  n^2-2 \beta  n r-2 \beta  r\right)+2 r \sinh ^2(r) \left(\alpha  m \left(4 m^2-11 m+7\right) 2^{m+n} \left(\frac{{r_0} {\sech}^2(r) \coth ({r_0})}{r}\right)^{m+n}\right.\right.\right.\nonumber\\
&+&\left.\left.\left.4 r \tanh (r) \left(\alpha  (m-1)^2 m 2^{m+n} \left(\frac{{r_0} {\sech}^2(r) \coth ({r_0})}{r}\right)^{m+n}+\beta  n^3+2 \beta  n^2+\beta  n\right)+4 \beta  n^3\right.\right.\right.\nonumber\\
&+&\left.\left.\left.11 \beta  n^2+7 \beta  n\right)+\alpha  m \left(m^2+1\right) 2^{m+n} \sinh (2 r) \left(\frac{{r_0} {\sech}^2(r) \coth ({r_0})}{r}\right)^{m+n}\right)+4 r^2 \left(r\right.\right.
\nonumber\\
&\times& \left.\left.
\left(2 \sinh ^2(r) \left(\alpha  (m-1)^2 m 2^{m+n} \left(\frac{{r_0} {\sech}^2(r) \coth ({r_0})}{r}\right)^{m+n}+\beta  n^3+2 \beta  n^2+\beta  n\right)+\alpha  (m-1) m \right.\right.\right.\nonumber\\
&\times&  \left.\left.\left.\left(-2^{m+n}\right) \left(\frac{{r_0} {\sech}^2(r) \coth ({r_0})}{r}\right)^{m+n}+\beta  n^2+\beta  n\right)+3 \beta  n \sinh (r) \cosh (r)\right)+2 r \cosh ^2(r)\right. \nonumber\\
&\times&\left.\left(\alpha  (m-1)^2 m 2^{m+n} \left(\frac{{r_0} {\sech}^2(r) \coth ({r_0})}{r}\right)^{m+n}+\beta  n^3+2 \beta  n^2+\beta  n\right)\right)+{r_0}^2 (\sinh (2 r)\nonumber\\
&-&2 r) {\sech}^2(r) \coth ({r_0})\Bigg]\Bigg|
\end{eqnarray}
\begin{figure}
	\centering
	\subfigure[$\rho+p_r$]{\includegraphics[scale=.5]{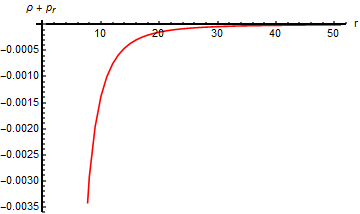}}\hspace{.5cm}
	\subfigure[$\rho-|p_r|$]{\includegraphics[scale=.5]{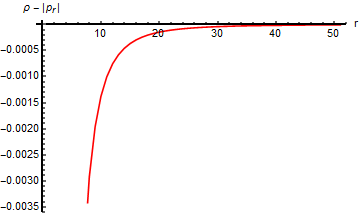}}\hspace{.5cm}
	\subfigure[$\triangle$]{\includegraphics[scale=.5]{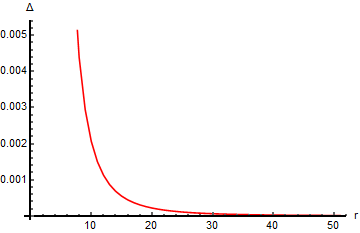}}\hspace{.5cm}
	\subfigure[$\omega$]{\includegraphics[scale=.5]{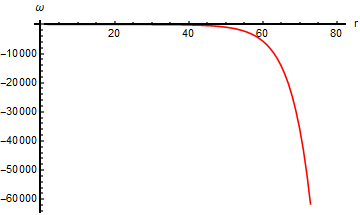}}\vspace{.5cm}
	\caption{Case 1: Plots for one NEC term, one DEC term, $\triangle$ \& $\omega$ with $f(R)=R$}
\end{figure}
\begin{figure}
	\centering
	\subfigure[$\rho$]{\includegraphics[scale=.46]{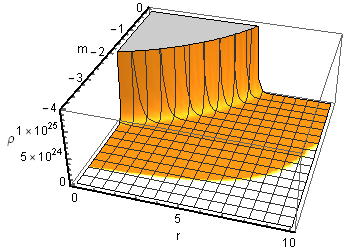}}\hspace{.1cm}
	\subfigure[$\rho+p_r$]{\includegraphics[scale=.46]{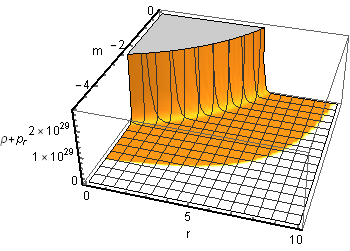}}\hspace{.1cm}
	\subfigure[$\rho+p_t$]{\includegraphics[scale=.46]{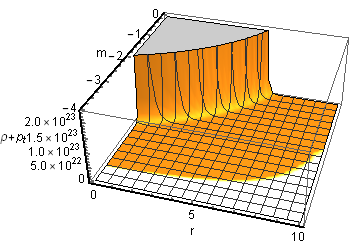}}\hspace{.1cm}
	\subfigure[$\rho-|p_r|$]{\includegraphics[scale=.46]{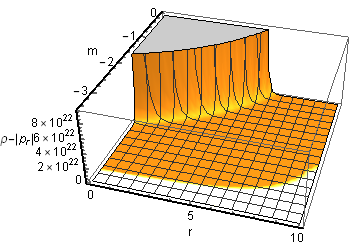}}\hspace{.1cm}
	\subfigure[$\rho-|p_t|$]{\includegraphics[scale=.46]{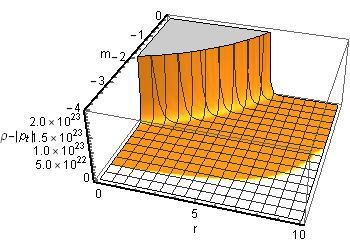}}\hspace{.1cm}
	\subfigure[$\triangle$]{\includegraphics[scale=.46]{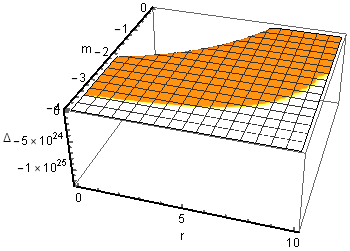}}\hspace{.1cm}
	\subfigure[$\omega$]{\includegraphics[scale=.46]{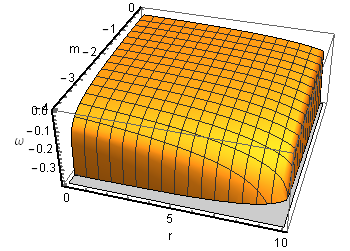}}\hspace{.1cm}
	\subfigure[$\rho+p_r$]{\includegraphics[scale=.45]{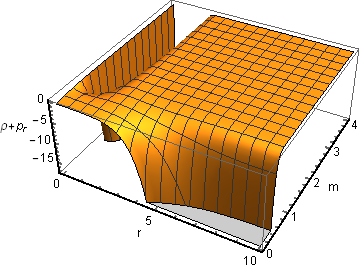}}\hspace{.1cm}
	\subfigure[$\rho-|p_r|$]{\includegraphics[scale=.45]{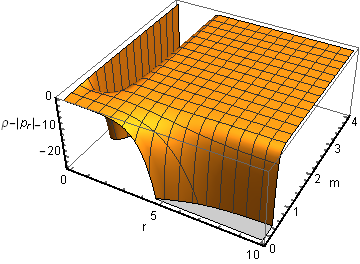}}\hspace{.1cm}
	\caption{Case 2: Plots for Density, NEC, DEC, $\triangle$ \& $\omega$ with $f(R) = R + \alpha R^m$}
\end{figure}

\begin{figure}
	\centering
	\subfigure[$\rho$]{\includegraphics[scale=.45]{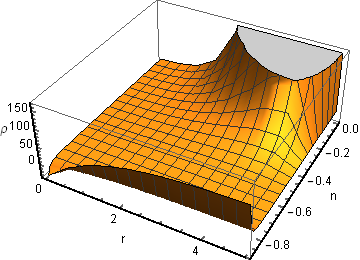}}\hspace{.01cm}
	\subfigure[$\rho+p_r$]{\includegraphics[scale=.45]{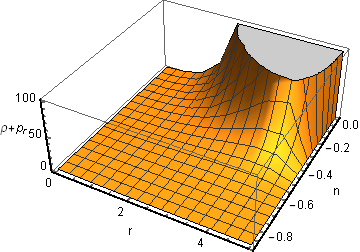}}\hspace{.01cm}
	\subfigure[$\rho+p_t$]{\includegraphics[scale=.45]{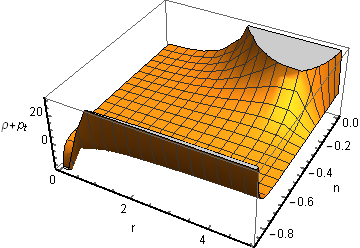}}\hspace{.01cm}
	\subfigure[$\rho-|p_r|$]{\includegraphics[scale=.45]{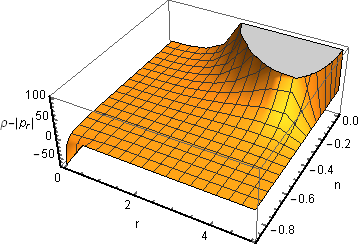}}\hspace{.01cm}
	\subfigure[$\rho-|p_t|$]{\includegraphics[scale=.45]{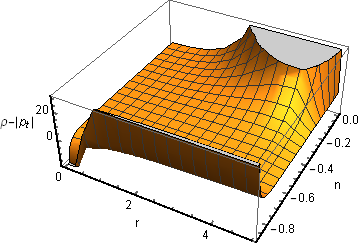}}\hspace{.01cm}
	\subfigure[$\triangle$]{\includegraphics[scale=.45]{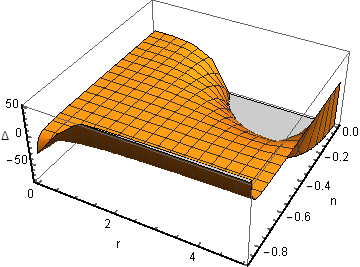}}\hspace{.01cm}
	\subfigure[$\omega$]{\includegraphics[scale=.45]{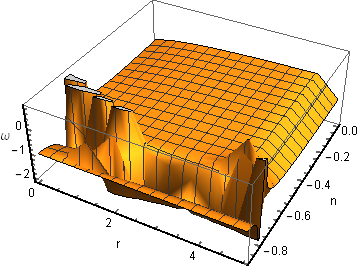}}\hspace{.01cm}
	\subfigure[$\rho+p_r$]{\includegraphics[scale=.45]{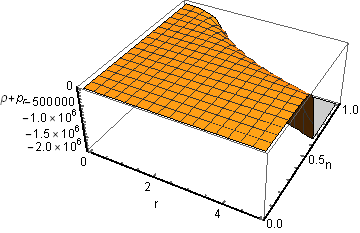}}\hspace{.01cm}
	\subfigure[$\rho-|p_r|$]{\includegraphics[scale=.46]{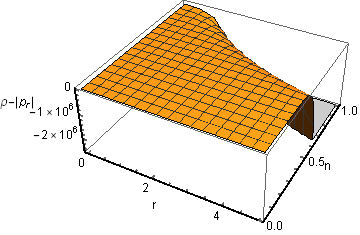}}\hspace{.01cm}
	\caption{Case 3: Plots for Density, NEC, DEC, $\triangle$ \& $\omega$ with $f(R)=R-\beta R^{-n}$}
\end{figure}
\begin{figure}
	\centering
	\subfigure[$\rho$]{\includegraphics[scale=.45]{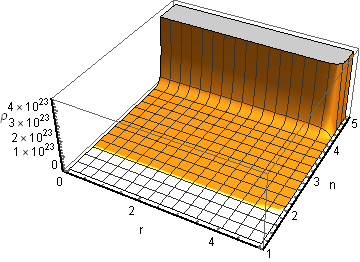}}\hspace{.01cm}
	\subfigure[$\rho+p_r$]{\includegraphics[scale=.45]{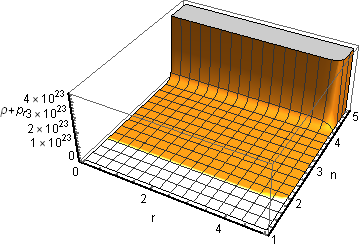}}\hspace{.01cm}
	\subfigure[$\rho+p_t$]{\includegraphics[scale=.45]{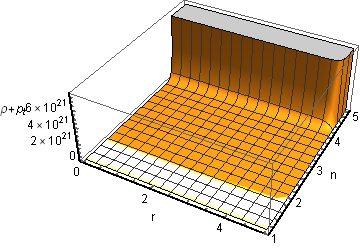}}\hspace{.01cm}
	\subfigure[$\rho-|p_r|$]{\includegraphics[scale=.45]{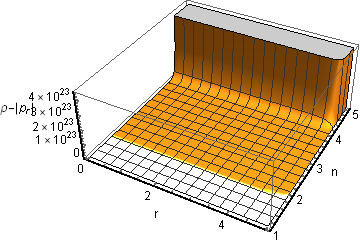}}\hspace{.01cm}
	\subfigure[$\rho-|p_t|$]{\includegraphics[scale=.45]{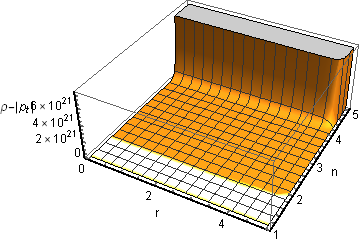}}\hspace{.01cm}
	\subfigure[$\triangle$]{\includegraphics[scale=.45]{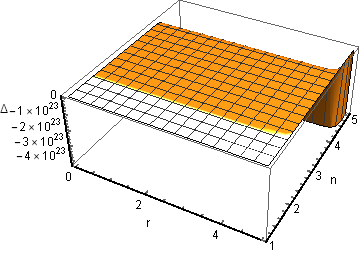}}\hspace{.01cm}
	\subfigure[$\omega$]{\includegraphics[scale=.45]{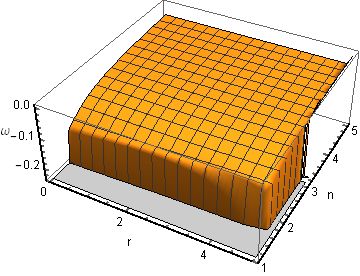}}\hspace{.01cm}
	\subfigure[$\rho+p_r$]{\includegraphics[scale=.45]{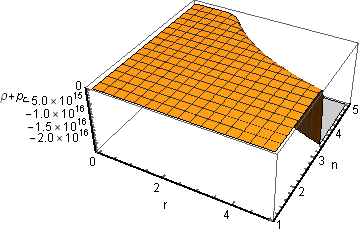}}\hspace{.01cm}
	\subfigure[$\rho-|p_r|$]{\includegraphics[scale=.45]{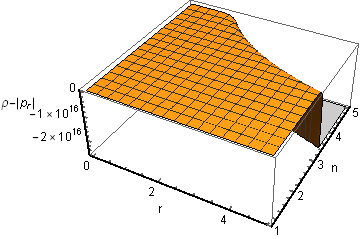}}\hspace{.01cm}
	\caption{Case 4: Plots for Density, NEC, DEC, $\triangle$ \& $\omega$ with $f(R)=R + \alpha R^m-\beta R^{-n}$}
\end{figure}

\section{Results \& Discussion}
The geometrical properties of wormholes are dependent on the shape function.  In literature, various shape functions are defined and wormhole structures are analyzed. The null energy condition (NEC) is defined for any null vector as
$NEC\Leftrightarrow T_{\mu\nu}k^{\mu}k^{\nu}\ge 0$, in terms of the principal pressures $NEC\Leftrightarrow ~~ \forall i, ~\rho+p_{i}\ge 0$. So, in this case we need to check $\rho+p_r$ and $\rho+p_t$. The weak energy condition (WEC) is defined for any timelike vector as $WEC\Leftrightarrow T_{\mu\nu}V^{\mu}V^{\nu}\ge 0$. WEC says that the energy density measured by any timelike observer should be positive locally. Therefore, it is physically important. In terms of the principal pressures $WEC\Leftrightarrow \rho\ge 0;$ and $\forall i,  ~~ \rho+p_{i}\ge 0$. So, we will examine $\rho$, $\rho+p_r$ and $\rho+p_t$.
The dominant energy condition (DEC) is defined for any time like vector $DEC\Leftrightarrow T_{\mu\nu}V^{\mu}V^{\nu}\ge 0$, and
$T_{\mu\nu}V^{\mu}$ is not space like. This says that the energy density is always positive locally, and that the energy flux is time-like or null. The dominant energy condition implies the null energy condition, but does not necessarily imply the strong energy condition. In terms of the principal pressures
$DEC\Leftrightarrow \rho\ge 0;$ and $\forall i, ~ p_i\in [-\rho, ~+\rho]$. For the model undertaken, we need to analyze the terms $\rho-|p_r|$ \& $\rho-|p_t|$.
In this paper, the shape function $ b(r)=\frac{r_0 \tanh(r)}{\tanh(r_0)}$ is chosen to explore the wormhole solutions in the background of $f(R)$ gravity with the function $f(R)=R+\alpha R^m-\beta R^{-n}$, where $\alpha$, $\beta$, $m$ \& $n$ are real numbers.
Different cases are considered according to the values of the parameters in  function $f(R)$ and in each case, null, weak \& dominated energy conditions, equation of state  and anisotropy parameters  are calculated.\\

\textbf{Case-1:} $\alpha=0$ \& $\beta=0$, i.e.  $f(R)=R$, general relativity case. In this case, the energy density $\rho$ is found to be a positive function of radial coordinate. The first null energy  condition term $\rho + p_r$ is found to be negative, while second null energy condition term $\rho + p_t$ is obtained to be positive.  The dominated energy condition terms $\rho-|p_r|$ \& $\rho-|p_t|$ are also obtained to be negative. This shows that the null, weak and dominated energy conditions are violated (Figs. 1(a), 1(b)). Hence, it is confirmed that  the wormhole is filled with exotic matter.  The anisotropy parameter ($\triangle$) is come out be a positive function of $r$ (Fig. 1(c)) This indicates that the geometry is repulsive. The equation of state parameter ($\omega$) is found to possess values less than -1 for every $r$ which shows the presence of phantom fluid (Fig. 1(d)). Hence, we could say that the presence of exotic matter is a necessary condition for the existence of wormhole solutions in general relativity with hyperbolic shape function. \\

\textbf{Case-2:} $\beta = 0$, i.e. $f(R) = R + \alpha R^m$, where $\alpha$ and $m$ are real numbers. First, the results are computed for  $\alpha>0$ with the variation of $m$. For $m\in(-\infty,0]\cup[2.2,\infty)$, the energy density is obtained to be positive with the variation of radial coordinate $r$ and else it is negative. The first null energy condition term $\rho+p_r$ is obtained to be positive for $m<0$ and negative for $m\geq0$. The second null energy condition term $\rho+p_t$ is obtained to be positive for $m\in (-\infty,0]\cup[1,\infty)$ and negative for $m\in (0,1)$. Further, the first dominated energy condition term is positive for $m<0$ and negative for $m\geq 0$. The second dominated energy condition term is positive for $m\leq0$ and negative for $m> 0$. Thus,  null, weak and dominated energy conditions are satisfied throughout for $m<0$ (Figs. 2(a)-2(e)), whereas the all energy conditions are violated for $m\ge 0$ (Figs. 2(h), 2(i)). Secondly, we obtained the results for $\alpha<0$ with the variation of $m$. The energy density is found to be positive with the increment of $r$ for $m\in (0,\infty)-\{1\}$. The terms $\rho+p_r$, $\rho-|p_r|$ and $\rho-|p_t|$ are obtained to be positive for $m\in(0,1)$ and $\rho+p_t$ is found positive for $m>0$. throughout. Hence, the model   $f(R) = R + \alpha R^m$ with (i) $\alpha>0$, $m<0$ \& (ii) $\alpha<0$, $m\in(0,1)$ confirms the existence of wormhole geometry without violating NEC, WEC and DEC. Therefore, we conclude that the non-exotic matter could support the existence of wormhole solutions in particular $f(R)=R + \alpha R^m$ model for (i) $\alpha>0$, $m<0$ \& (ii) $\alpha<0$, $m\in(0,1)$. However, the all energy conditions are violated otherwise. Hence the parameter $m$ play as an important role in this particular model for violation and non-violation of energy conditions.
Subsequently, for $\alpha>0$, the anisotropy parameter is observed to be negative for $m<0$ (Fig. 2(f)) and positive for $m\geq 0$. For $\alpha<0$, it is negative for $m\in(0,1)$ and positive for $m\in (-\infty,0]\cup[1,\infty)$.
Further, for $\alpha>0$, the equation of state parameter $\omega$ lies between -1 and 0 (Fig. 2(g)) and $\omega$ is positive for $m\leq 1$ and $1<m\leq 2$ respectively, which indicates that the presence of non-phantom fluid in wormhole geometry for $m<0$. For $\alpha<0$, $-1<\omega<0$ and $\omega<-1$  for $m\in(0,1)$ and $m\in (-\infty,0]\cup[1,\infty)$ respectively, which indicates that the presence of non-phantom fluid in wormhole geometry for $m\in(0,1)$.  However, $\omega<-1$ for (i) $\alpha>0$, $m\ge 2$ \& (ii) $\alpha<0$, $m\in(0,1)$ which indicates that the presence of phantom fluid in wormhole geometry.
%Therefore, all energy conditions are violated for $m\ge 0$. \\
\\

\textbf{Case-3:} $\alpha=0$, i.e. $f(R)=R-\beta R^{-n}$, where $\beta$ and $n$ are real numbers. Let $\beta>0$ and $n$ be any real number. Then, the energy density is found to be a positive function of $r$ for $n<0$. However,  the energy density is found to be negative for $n>0$, which indicates the presence of exotic matter. The first null energy condition term is positive for every $r>0$ when $-1< n \leq 0$ and else it is negative. The second null energy condition term is positive for $r>2.6$ \& $n\leq 0$ and negative otherwise.  Both first and second dominated energy condition terms are obtained to be positive for every $r>0$ and $-1< n \leq 0$. Thus, null, weak and dominated energy conditions are satisfied for $-1< n\leq 0$ (Figs. 3(a)-3(e)) and violated for $n\leq-1$ or $n>0$ (Figs. 3(h)-3(i)). The anisotropy parameter is obtained to be positive for $n\leq-1$ and negative for $n> -1$ (Fig. 3(f)). The  equation of state parameter is found to be less than -1 for $n\leq-1$ and lies between -1 and 0 for $n>-1$ (Fig. 3(g)). Thus, in this case, for $-1< n\leq 0$, wormhole solutions with attractive geometry exist without any violation of energy conditions.  Now, we assume that $\beta<0$ and $n$ is any real number. Then $\rho$, $\rho+p_t$ \& $\rho-|p_t|$ are positive for $n\geq0$ and $\rho+p_r$ \& $\rho-|p_r|$ are positive for $n>0$. Thus, null, weak and dominated energy conditions are satisfied for $n>0$ and violated for $n\leq 0$. Now, for $n>0$, $\triangle<0$ and for $n\leq 0$, $\triangle>0$. For $n>0$, $-1<\omega<0$ and for $n\geq 0$, $\omega<-1$. Hence, the wormhole structure with attractive geometry exists without presence of exotic matter and phantom fluid for $\beta<0$ and $n>0$. Consequently, the model $f(R)=R-\beta R^{-n}$ is significant with (i) $\beta>0$, $-1< n\leq 0$ \& (ii) $\beta<0$, $n>0$.   \\

\textbf{Case-4:} $\alpha\neq0$ \& $\beta\neq0$, i.e. $f(R)=R+\alpha R^m-\beta R^{-n}$.

Subcase (i): $\alpha>0$, $\beta>0$. In this subcase, the energy density and null \& dominated energy condition terms are positive if $m<0$ \& $n<-m$. In this  range of $m$ and $n$, $\triangle<0$  and $-1<\omega<0$.

Subcase (ii): $\alpha>0$, $\beta<0$. In this subcase, the energy density and null \& dominated energy condition terms are positive under the following ranges of parameters $m$ \& $n$: (a) $m>0$, $n>0$, (b) $m\leq -1$, $n\geq 0$ and (c) $-1<m<0$, $\forall n$. For these ranges $\triangle<0$  and $-1<\omega<0$.

Subcase (iii): $\alpha<0$, $\beta>0$. In this subcase, the energy density and null \& dominated energy condition terms are positive under the following ranges of parameters $m$ \& $n$: (a) $0<m<1$, $n\leq0$ and (b) $m\geq1$, $-1<n<0$. For these ranges also $\triangle<0$  and $-1<\omega<0$.
	
Subcase (iv): $\alpha<0$, $\beta<0$. In this subcase, the energy density and null \& dominated energy condition terms are positive under the following
ranges of parameters $m$ \& $n$: (a) $m=0$, $n>0$, (b) $m\geq -1$, $n> 0$, (c) $0<m<1$, $n<-m$, (d) $0<m<1$, $n\geq 0$ and (e) $m<0$, $n>-m$. For these ranges $\triangle<0$  and $-1< \omega<0$.	In this case, the results are plotted only for subcase (i) taking $m=-5$. Figs. (4(a)-4(g)) show the satisfaction of energy conditions and presence of attractive geometry with non-phantom fluid. However, in Figs. (4(h),4(i)) the violation of energy conditions is shown taking $m$ positive, in particular $m=5$ is taken here.

Thus, we analyze that the results obtained in Cases 2-4 are completely different from those obtained in Case 1.
For each case, we are interested in the ranges of parameters where energy conditions are satisfied. These ranges are summarized in the table below:
\begin{table}[!h]
	\centering
	\caption{Range of parameters in $f(R)=R+\alpha R^m-\beta R^{-n}$,  where NEC, WEC \& DEC are satisfied}
	\begin{tabular}{|c|l|}
		\hline
		\textbf{S. No.}& \textbf{Range  of parameters} \\
		\hhline{|=|=|}
       1. &$\alpha>0$, $\beta=0$, $m<0$,  $\forall n$\\\hline
       2. &$\alpha<0$, $\beta=0$, $m\in(0,1)$,  $\forall n$\\\hline
    3.&$\alpha=0$, $\beta>0$, $\forall m$, $-1< n \leq 0$  \\\hline
    4.&$\alpha=0$, $ \beta<0$, $\forall m$, $n>0$  \\\hline

%		 \multirow{11}{*}{$f(R)=R+\alpha R^m-\beta R^{-n}$}
	5.	                &  $ \alpha>0$, $\beta>0$,  $m<0$, $n<-m$\\\hline
	6.				   &  $\alpha>0$, $\beta<0$, $m>0$, $n>0$\\\hline
					  7. &  $\alpha>0$, $\beta<0$, $m\leq -1$, $n\geq 0$\\\hline
					  8.& $\alpha>0$, $\beta<0$, $-1<m<0$, $\forall n$\\\hline
					  9. &  $\alpha<0$, $\beta>0$, $0<m<1$, $n\leq0$ \\\hline
					   10.& $\alpha<0$, $\beta>0$, $m\geq1$, $-1<n<0$\\\hline
					   11. & $\alpha<0$, $\beta<0$, $m=0$, $n>0$\\\hline
					  12.& $\alpha<0$, $\beta<0$, $m\geq 1$, $n> 0$\\\hline
					   13.& $\alpha<0$, $\beta<0$, $0<m<1$, $n<-m$\\\hline
					   14.& $\alpha<0$, $\beta<0$, $0<m<1$, $n\geq 0$ \\\hline
			           15.& $\alpha<0$, $\beta<0$, $m<0$, $n>-m$\\\hline
			\end{tabular}
\end{table}

\section{Conclusion}
The modified gravity which naturally merges  two expansion phases of the Universe: inflation
at early times (rapid expansion) and cosmic acceleration at the current
epoch. The higher derivative terms with positive power of
curvature are dominant at the early universe providing the
inflationary stage. The terms with negative power of curvature
serve as a gravitational alternative for dark energy making
possible the cosmic speed-up.  Moreover, it may pass the simplest solar system constraint for scalar-tensor gravity equivalent
to original modified gravity. Clearly, more investigation of this
theory should be needed in order to conclude whether the model is a realistic one or not.

In this paper, static traversable wormholes are investigated by using a hyperbolic shape function in terms  of radial coordinate. The Einstein's field equations derived in the framework of $f(R)$ gravity are used to obtain the wormhole solutions. To find the exact solution, the well known form of the function $f(R)$ defined as $f(R) = R + \alpha R^m - \beta R^{-n}$ is taken\cite{noj1} into account. Depending on the values of $\alpha$, $\beta$, $m$ and $n$, the energy conditions, anisotropic parameter and equation of state  parameter are determined in four cases. It is found that the energy conditions, namely null energy condition, weak energy condition and dominated energy condition are satisfied  in each case except GR case. In GR case, geometry is repulsive and filled with phantom fluid, while in all other cases of $f(R)$ gravity, the geometric nature is attractive and filled with like non-phantom fluid for some range of $\alpha$, $\beta$, $m$ and $n$.   Thus, the existence of wormhole solutions is possible without violation of energy conditions using $f(R)$ theory of gravity. This represents the significance of the framework of modified theory of gravity. Thus, all the conditions for the existence of wormhole geometries are fully satisfied and strongly confirm their presence in the universe.\\

\textbf{Acknowledgement:} Authors are very much thankful to the anonymous reviewer for his
constructive comments for improvement of the manuscript. The second author G. C.
Samanta is extremely thankful to Council of Scientific and Industrial Research (CSIR), Govt. of
India, for providing financial support (Ref. No. 25(0260)/17/EMR-II) to carrying out the
research work.

\end{document}